\documentclass[%
	reprint,
	amsmath,amssymb,
	aps,
    pre,
    superscriptaddress
]{revtex4-2}

\usepackage{layout}
\usepackage{graphicx}
\usepackage{caption}
\usepackage{subcaption}
\usepackage{dcolumn}
\usepackage{bm}
\usepackage[english]{babel}
\usepackage{xcolor}
\usepackage{bbold}

\makeatletter
\let\l@en\l@english
\makeatother

\usepackage[colorlinks,
	pdffitwindow=false,
	plainpages=false,
	pdfpagelabels=true,
	pdfpagemode=UseOutlines,
	pdfpagelayout=SinglePage,
	bookmarks=false,
	colorlinks=true,
	hyperfootnotes=false,
	linkcolor=blue,
	urlcolor=blue!50!black,
	citecolor=green!50!black,
    breaklinks]
{hyperref}

\newtheorem{rem}{Remark}
\usepackage{cleveref}

\crefname{figure}{Fig.}{Figs.}
\Crefname{figure}{Figure}{Figures}
\crefname{equation}{Eq.}{Eqs.}
\Crefname{equation}{Equation}{Equations}
\crefname{appendix}{Appendix}{Appendices}
\Crefname{appendix}{Appendix}{Appendices}

\newcommand{\R}{\mathbb{R}}

\renewcommand{\P}{\mathbb{P}}
\newcommand{\E}{\mathbb{E}}
\newcommand{\A}{\mathbb{A}}
\newcommand{\M}{\mathbb{M}}

\newcommand{\dif}{\mathrm{d}}

\newcommand{\rv}[1]{\bm{#1}}  

\begin{document}

\title{Collective variables for homophily-driven network rewiring dynamics}

\author{Sören Nagel}
\affiliation{Zuse Institute Berlin, Germany}%
\affiliation{Freie Universit\"at Berlin, Germany}
\author{Stefanie Winkelmann}
\affiliation{Zuse Institute Berlin, Germany}%
\author{Péter Koltai}
\affiliation{Department of Mathematics, University of Bayreuth, Germany}%
\author{Nataša Djurdjevac Conrad}
\affiliation{Zuse Institute Berlin, Germany}%
\author{Marvin Lücke}
\affiliation{Zuse Institute Berlin, Germany}%

\begin{abstract}
Stochastic network rewiring processes, in which edges dynamically rewire based on fixed node attributes, are widely used in applications ranging from social dynamics to neuroscience and form an important component of adaptive network modelling. 
In this paper, we identify low-dimensional collective variables (CVs) that capture the essential macroscopic behavior of such time-evolving networks and enable reduced-order descriptions of their dynamics.
To this end, we apply the data-driven transition manifold approach to homophily-driven rewiring models, in which edges preferentially connect nodes with similar attributes.
For two representative models, we find that the optimal CV is a consensus measure quantifying the fraction of edges whose incident nodes differ by less than a certain threshold.
Building on the learned CV, we construct reduced macroscopic models using a data-driven approach based on sparse regression and through an analytical derivation using graphons. The latter yields a closed-form evolution equation for the consensus measure and analytically validates the identified CV.
\end{abstract}

\maketitle

\section{\label{sec:Intro}Introduction}
Time-evolving networks defined by rewiring processes are a central component of adaptive (or co-evolving) network models, which arise in diverse applications such as opinion dynamics, epidemiology, neurodynamics, and ecology~\cite{berner_adaptive_2023, sayama_modeling_2013, zschaler_adaptive-network_2012}.
In the adaptive network models considered in these works, each node of the network is endowed with a state that evolves over time depending on the states of the other nodes and on their connections.
In this context, nodes are also often referred to as \textit{agents}.
The network itself changes over time by edge rewiring or even by adding or deleting nodes, depending on the nodes' states.
These adaptive network models are, for instance, commonly employed in models of opinion dynamics \cite{kan_adaptive_2023, kozma_consensus_2008, vazquez_generic_2008, peng_adaptive_2025, su_coevolution_2014, maia_adaptive_2021, holme_nonequilibrium_2006}, where the state of an agent represents its opinion on a given issue, and the network describes social interactions or relations between the agents.
Beyond opinion dynamics, adaptive network models also arise in many other application areas. In epidemiology, for example, nodes may represent individuals and edges represent physical contacts through which infections spread, while in neurodynamics or ecology, they can describe interacting neurons or species \cite{PhysRevLett.96.208701, topping_understanding_2021, rubinov_symbiotic_2009, Staniczenko2010, fialkowski_heterogeneous_2023, PhysRevLett.134.047402}.

In this work, we focus on the network rewiring aspect of such adaptive models, i.e., the edges in the network change over time while node states remain static.
This reduces the dimensionality of the state space and provides a natural starting point for algorithmically learning the collective behavior of more general co-evolving network models \cite{djurdjevac_conrad_co-evolving_2024}.
Moreover, many adaptive network models exhibit a separation of timescales, so the analysis presented here can be interpreted as studying the regime in which the graph evolves much faster than the node states.
This is, for instance, the case in certain infection-spreading models, where the network represents physical contacts between individuals, which may change daily, while their infection status can remain unchanged for weeks or months \cite{gubela_behavior_2026}.

Even though the microscopic rules governing node and edge behavior are often simple, the emerging macroscopic system behavior can be complex and hard to predict.
The objective of this work is to algorithmically learn the collective macroscopic behavior of time-evolving networks by finding a low-dimensional representation of the system that captures the fundamental dynamics on timescales of interest.
The mapping into this low-dimensional space defines a \textit{collective variable} (CV).
In the literature, a CV is considered \textit{good} if it is low-dimensional, interpretable in the respective modelling context, and retains the essential information about the system's behavior, i.e., allowing the definition of an approximate macroscopic model that closely replicates the projection of the original system into CV space~\cite{rogal2021reaction,lucke_concentration_2024}.
We apply the \textit{transition manifold approach} \cite{bittracher_transition_2018, lucke_learning_2024} to learn good CVs based on simulation data of time-evolving network models.
Furthermore, we show how to use the resulting CVs to obtain a macroscopic system approximation, either by appending another data-driven method for learning the reduced dynamics in the low-dimensional CV space or by using the CVs as a foundation for analytical derivation.

Specifically, we focus on stochastic rewiring processes driven by (opinion) homophily \cite{khanam_homophily_2023}, i.e., given the static state of each node, edges are more likely to rewire between nodes with similar states.
We discuss two examples of such models and show that for both models the optimal CV is a simple homophily (or consensus) measure that counts the number of edges for which the state difference of adjacent nodes is smaller than a certain threshold.
We find that other commonly used measures for consensus or polarization, such as the attribute assortativity~\cite{kan_adaptive_2023}, are less suitable for defining the macroscopic dynamics.
This insight may also be valuable when analyzing similar opinion dynamics models in future work.
Moreover, we demonstrate that, as a next step, a macroscopic system can be learned from data using methods such as SINDy~\cite{Brunton2016}, which yields a simple ODE that describes how the consensus evolves over time.
For one of the models, we derive a graphon~\cite{Bayraktar2023} approximation based on the learned CV to obtain a macroscopic approximation in the large-population limit.
For both the data-driven SINDy approach and the analytical graphon approach, we verify that the derived macroscopic system provides an accurate approximation of the projected microscopic dynamics, which in turn validates the identified~CV.


\section{\label{sec:Models} Model Setup}
In this paper, the terms ``network'' and ``graph'' will be used synonymously.
Let $G = (V, E)$ be a simple (undirected, unweighted) graph with node set $V = \{1, \dots, N\}$.
We denote an edge between two nodes $i,j \in V$ as $(i,j)$.
Since the graph is undirected, $(i,j)$ and $(j,i)$ are equivalent.
We assume that $V$ is fixed over time, i.e., no nodes are added or removed. However, the edges $E$ change over time due to a stochastic rewiring process detailed later.
This defines the graph-valued stochastic process $\rv{G}(t) = (V, \rv{E}(t))$.
In the following, we represent a graph $G$ by its symmetric adjacency matrix $A \in \{0, 1\}^{N \times N}$, and denote the corresponding stochastic process by $\rv{A}(t)$.
Furthermore, each node $i \in V$ is endowed with a state $\theta_i \in [0, 1]$ that is static over time.
Without loss of generality, we assume that the nodes are labeled in ascending order of their states, i.e., $i \leq j$ implies $\theta_i \leq \theta_j$.
We interpret this system in the context of opinion dynamics, and hence refer to a node $i$ as an \textit{agent} and to its state $\theta_i$ as its \textit{opinion}.
We will later consider several sets of opinions $(\theta_1, \dots, \theta_N)$ sampled from different distributions to demonstrate the generality of our results.

A key mechanism driving opinion formation in networks is \textit{homophily}, i.e., the tendency of agents to interact with others holding similar opinions
\cite{khanam_homophily_2023}.
In the models considered here, homophily is incorporated through the rewiring probabilities, which depend on the opinion similarity between agents. More precisely, we define the \textit{rewiring probability} $p_{i \to j}(A)$ as the probability that a node $i \in V$
rewires to a node $j \in V$ that is not currently its neighbor, given the adjacency matrix~$A$.
These rewiring probabilities play a central role in the edge update steps of the systems, which will be discussed in detail later.
We assume that $p_{i \to j}(A)$ increases with the opinion similarity $1 - |\theta_i - \theta_j|$, more precisely,
\begin{equation}
	p_{i \to j}(A) := \frac{1}{Z_i(A)} (1 - A_{ij})(1- |\theta_i - \theta_{j}|), \label{eq:rewiring_probability}
\end{equation}
with normalization constant
\begin{equation}
	Z_i(A) := \sum_{k} (1 -  A_{ik}) (1- |\theta_i - \theta_{k}|).
\end{equation}

In the following, we describe two models inspired by the rewiring mechanism of the adaptive model presented in \cite{kan_adaptive_2023}, namely the \textit{ergodic model} and the \textit{threshold model}, which differ in their rewiring rules.
In both models, the total number of edges
\begin{equation}
K := |E| = \sum_{i<j}A_{ij}
\end{equation}
is constant over time because, in each rewiring step, one edge is added and one edge is removed.
Therefore, given a fixed number of edges $K$, the stochastic process $\rv{A}(t)$ takes values in the set of adjacency matrices
\begin{equation*}
	\A := \left\{ A \in \{0,1\}^{N \times N} \,\bigg\vert\, A^T = A; A_{ii}=0; \sum_{i<j}A_{ij} = K\right\}.
\end{equation*}
The methods introduced later could also be adapted to dynamics with a varying number of edges, but in this work we focus on a fixed number of edges.

\subsection{Ergodic Model}\label{sec:ergodic_model}
The transition manifold approach, which we apply later, yields optimal collective variables for reversible and ergodic stochastic systems with a unique invariant measure~\cite{bittracher_transition_2018}.
Therefore, we first consider an ergodic rewiring model adapted from \cite{kan_adaptive_2023}.
The edge update consists of the following steps:
\begin{enumerate}
	\item Pick a node $i$ uniformly at random.
	\item Select a neighbor $j$ of node $i$ uniformly at random.
	\item Sample a potential new neighbor $j^*$ according to the rewiring probabilities $p_{i \to j^*}$ defined in~\Cref{eq:rewiring_probability}.
	\item If $j^*$ is closer to $i$ in opinion than $j$, i.e., $|\theta_i - \theta_{j^*}| \leq |\theta_i - \theta_j|$, the edge $(i, j^*)$ replaces the edge $(i, j)$.
    Otherwise, this replacement is performed anyway with probability~$p\in(0,1)$.
\end{enumerate}
We model the rewiring dynamics in continuous-time from an agent-based perspective.
We assume that each node executes steps~2--4 after exponentially distributed waiting times independently of the others, leading naturally to a Poisson process description.
Specifically, we define the stochastic system $\rv{A}(t)$ by performing the above edge update at the event times of a global Poisson process with rate $\lambda>0$.
The parameter $\lambda$ represents the total update rate.
Assuming constant individual update rates, this implies the scaling $\lambda \sim N$ such that the expected number of updates per agent per unit time remains constant.
Furthermore, modelling the dynamics in continuous-time allows evaluation at arbitrary time points.

\begin{figure}
	\centering
	\includegraphics[width=\linewidth]{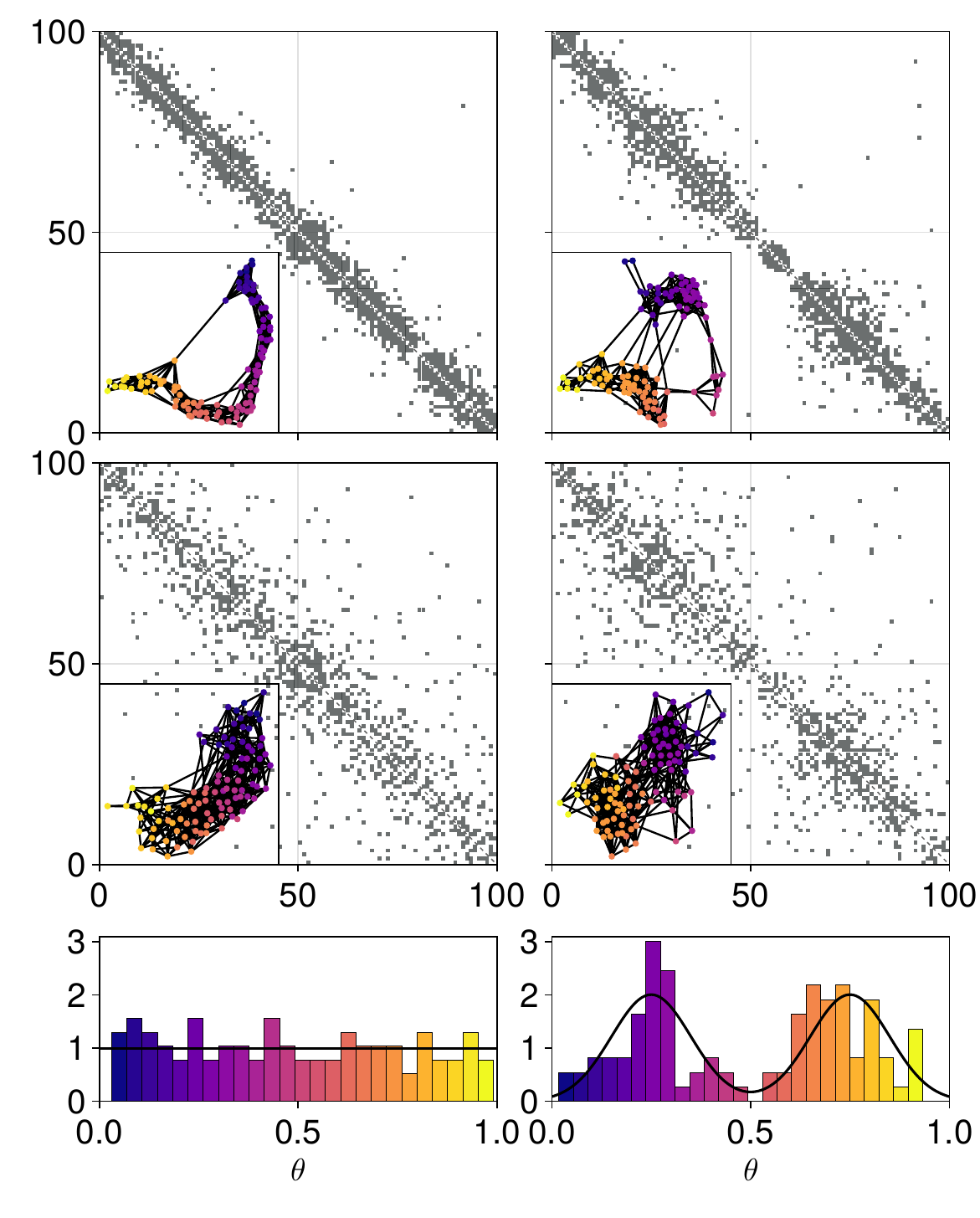}
	\caption{Illustration of the process state $A(t)$ (adjacency matrix) at time $t=150$ for exemplary realizations of the ergodic model with different parameters.
    The inlays display the corresponding network, where node colors indicate the opinion values $(\theta_1,\dots,\theta_N)$.
    Opinions are drawn from a uniform distribution (left column), or a bimodal distribution (right column).
    Results are shown for acceptance probabilities $p=0.01$ (top row) and $p=0.1$ (center row). Other parameters: $N=100$, $K=500$, $\lambda = N$.}
	\label{fig:smooth_model}
\end{figure}

In \Cref{fig:smooth_model}, we show examples of the process by plotting the adjacency matrix $\rv{A}(t=150)$ for different values of $p$ and opinion distributions.
The opinion distribution determines the clustering of the networks, i.e., for uniformly distributed opinions the adjacency matrix has a more banded structure, whereas a bimodal distribution results in the formation of two main clusters. 
The acceptance probability $p$ acts as a noise (or temperature) parameter: For $p\to 0$, any new edge must connect nodes of more similar opinions than the old edge; for $p\to 1$, it becomes more likely that a node rewires to another with more dissimilar opinion.

\subsection{Threshold Model} \label{sec:threshold_model}
In addition to homophily, another key concept in opinion dynamics is \textit{bounded confidence}~\cite{rainer2002opinion}, which restricts interactions to agents with sufficiently similar opinions.
This is commonly implemented by requiring the opinion difference between interacting agents to remain below a threshold~$r>0$.
Here, we adopt the specific rewiring mechanism from the adaptive model presented in \cite{kan_adaptive_2023}.

Edges are called \textit{concordant} if the opinion difference between their incident nodes is smaller than the threshold~$r$, i.e., the edge $(i, j)$ is concordant if $|\theta_i -  \theta_j| \leq r$.
Edges that are not concordant are called \textit{discordant}.

The edge update of the threshold model consists of the following steps:
\begin{enumerate}
	\item Select a discordant edge $(i, j)$ uniformly at random from all discordant edges.
	\item Pick one of the incident nodes $i$ or $j$ uniformly at random for rewiring. Without loss of generality, assume that node $j$ is selected.
	\item Sample a new neighbor $j^*$ to replace $j$ according to the rewiring probabilities $p_{i \to j^*}$ defined in~\Cref{eq:rewiring_probability}.
	\item Replace the edge $(i, j)$ with the edge $(i, j^*)$.
\end{enumerate}
Again, the continuous-time network dynamics $\rv{A}(t)$ is driven by a Poisson process with an event rate $\lambda$ by conducting the above edge update at each event.

Let $\A_c := \{A \in \A \mid \forall i,j: A_{ij} = 1 \Rightarrow |\theta_i - \theta_j| \leq r\}$ be the set of adjacency matrices containing only concordant edges.
Note that $\A_c$ is also the set of absorbing states of the dynamics, since if $\rv{A}(t) \in \A_c$, there are no discordant edges left to rewire.
Let $K_c$ denote the total number of possible concordant edges, $K_c := \sum_{i<j} \mathbb{1}_{[0, r]}(|\theta_i -  \theta_j|)$.
If $K \leq K_c$, i.e., the number of edges $K$ of the graph is less than or equal to the number of possible concordant edges $K_c$, then there exists at least one absorbing state, $\A_c \neq \emptyset$.
Hence, once $\rv{A}(t) \in \A_c$, the process remains in that state forever.
Consequently, the process is not ergodic and does not admit a unique invariant measure.
If, however, $K > K_c$, then there are no absorbing states, i.e., $\A_c = \emptyset$.
After all possible concordant edges have been formed, the remaining discordant edges continue rewiring indefinitely.
In this case, the system exhibits a unique invariant measure.

\begin{figure}
	\centering
	\includegraphics[width=\linewidth]{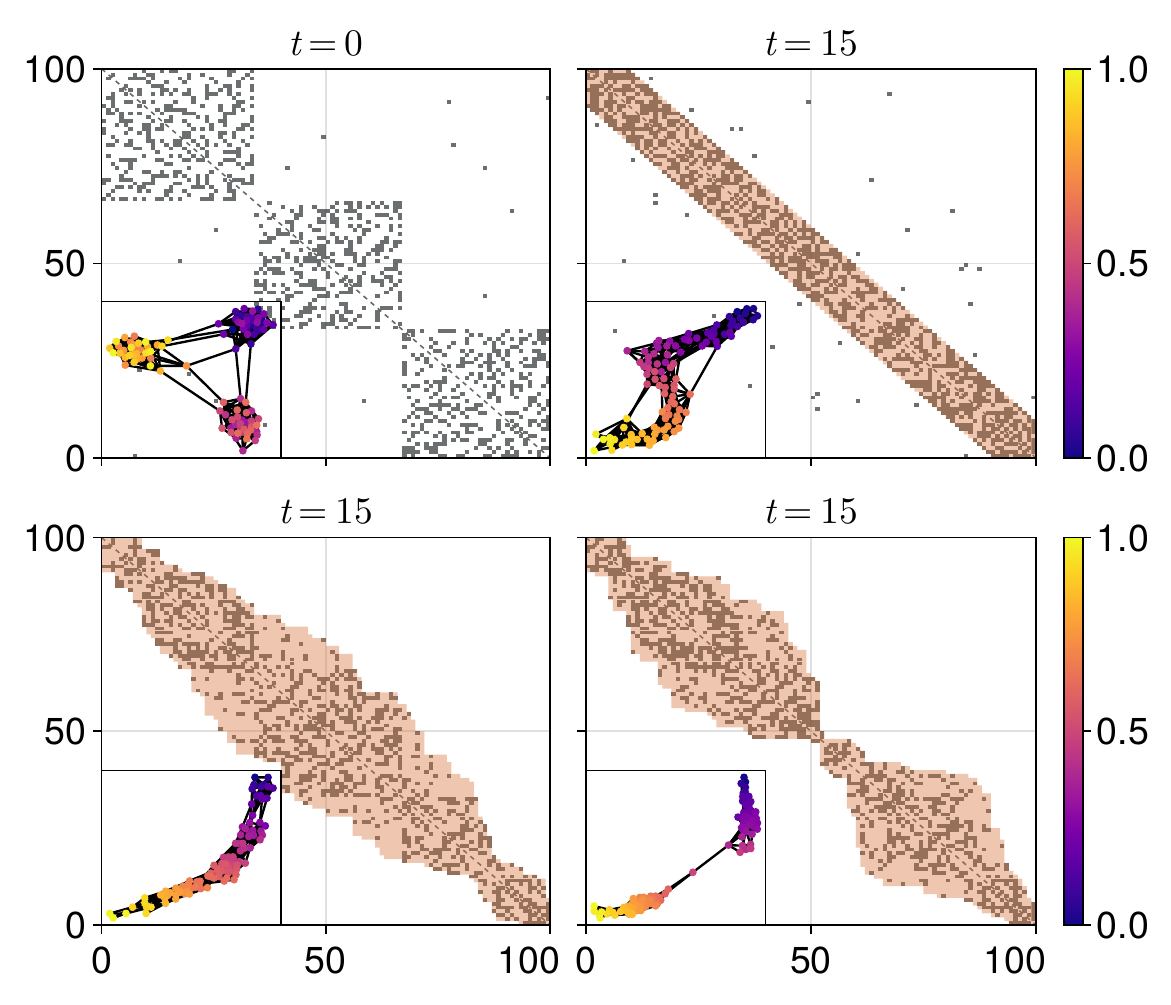}
	\caption{
	Illustration of the process state $\rv{A}(t)$ (adjacency matrix) at time $t=15$ for exemplary realizations of the threshold model with the same initial graph (top left) and opinions $(\theta_1,\dots,\theta_N)$ distributed in either of three ways: equidistant $\theta_{i+1} - \theta_i = 1/N$ (top right); or drawn from a normal distribution with mean $\mu=0.5$ and variance $\sigma^2 = 0.2$ (bottom left) as in~\cref{fig:smooth_model_cv}; or drawn from a bimodal distribution (bottom right) as in~\cref{fig:smooth_model}.
    The shaded region illustrates the so-called concordant set, i.e., if all edges of $A$ are in the shaded region then $A$ is an absorbing state, $A \in \A_c$.
	The inlays display the corresponding network, where node colors indicate the opinion values.
	Other parameters are $N=100$, $K=500$, $\lambda = N$, $r=0.1$.}
	\label{fig:threshold_model_dynamics}
\end{figure}

In \Cref{fig:threshold_model_dynamics} (all panels except top left), we show examples of the process by plotting the adjacency matrix $\rv{A}(t)$ for different opinion distributions and with $K < K_c$, so eventually all edges become concordant and lie within the orange shaded region, which marks the set of all possible concordant edges. The opinion distribution determines the shape of the concordant domain. This also has an effect on the timescale needed to reach an absorbing state. For the uniform opinion distribution the example has not yet reached such an absorbing state.

In the following sections, we show that the transition manifold approach yields meaningful collective variables and a corresponding macroscopic model even in the non-ergodic case.

\section{\label{sec:TMfd}Transition Manifold Approach}
As defined previously, let $\A$ denote the set of symmetric adjacency matrices corresponding to graphs with a fixed number $K$ of edges.
Note that this set contains $|\A| = \binom{N(N-1)/2}{K}$ matrices.
We study a continuous-time stochastic rewiring process $\rv{A}(t) \in \A$, which in this section is assumed to admit a unique stationary distribution.
For $A \in \A$ and a \textit{lag time} $\tau > 0$, let $p_A^\tau \in \R^{|\A|}$ denote the (discrete) probability distribution of the system at time $\tau$ after starting in $A$ at time $0$.
That is, for each $B \in \A$
\begin{equation}
	p_A^\tau(B) := \P(\rv{A}(\tau) = B \mid \rv{A}(0) = A).
\end{equation}
The \textit{transition manifold approach} \cite{bittracher_transition_2018} exploits the observation that for systems exhibiting a low-dimensional macroscopic approximation, and for an appropriate choice of $\tau$, the set
\begin{equation}
	\M_\tau := \{ p_A^\tau \mid A \in \A \} \subset \R^{|\A|}
\end{equation}
is well approximated by a $d$-dimensional submanifold $\M \subset \R^{|\A|}$, called the \textit{transition manifold}.
The lag time $\tau$ must be chosen appropriately for the system at hand, i.e., large enough so that fast subprocesses decay, but smaller than the time it takes to converge to the stationary distribution.
Assume that there exists a parametrization $\varphi: \M_\tau \to \R^d$, defined by first mapping each element of $\M_\tau$ to its best approximation on the $d$-dimensional manifold $\M$, and then to a representation of $\M$ in $\R^d$.
One can show that, if $\M_\tau$ is sufficiently close to $\M$ (see \cite{bittracher_transition_2018} for rigorous statements), the mapping $\varphi$ is a good \textit{collective variable} (CV). This means that there exist distributions $\smash{ p^\tau_{\varphi(A)} \in \R^{|\A|} }$ that depend only on the macroscopic information $\varphi(A)$, such that for all $A$
\begin{equation}
	p^\tau_A \approx p^\tau_{\varphi(A)}.
\end{equation}
In words, the essential information required to characterize the dynamics is captured by the collective variable~$\varphi$.

In practice, the transition manifold and the associated collective variable $\varphi$ can be approximated from simulation data of the system.
First, we choose a diverse set of adjacency matrices $A^1, \dots, A^L \in \A$ which we call \textit{anchor points}.
Here, diversity means that the anchor points cover the dynamically relevant region of $\A$ in which the system is likely to be found.
The anchor points are chosen appropriately if the associated distributions $p^\tau_{A^1}, \dots, p^\tau_{A^L}$ cover $\M_\tau$ sufficiently well.
Otherwise, the learned parametrization $\varphi$ may be biased due to insufficient coverage, leading to a CV that does not adequately characterize the entire process.
A general criterion for sufficient coverage cannot be given, since it depends on the specific system at hand.
For the models studied here, we generate anchor points using the following method:
\begin{enumerate}
	\item Sample $A \in \A$ uniformly at random using the Erd{\H o}s--R\'enyi random graph model.
	\item Sample a time $t$ uniformly at random from the interval $[0, T]$.
	\item Sample the anchor as the state $\rv{A}(t)$ obtained from the trajectory started at $\rv{A}(0) = A$.
\end{enumerate}
Here, $T$ should be chosen large enough so that, for any initial condition, $\rv{A}(T)$ has approximately reached the stationary distribution.
This method creates diverse anchor points since it uses an unbiased uniform sample of the state space $\A$ and then integrates the system forward to focus on dynamically relevant regions.
Since the integration time $t$ is also chosen randomly, we obtain samples from both the transient phase and the equilibrium.

In the next step of the transition manifold approach, we conduct $S\in\mathbb{N}$ short burst simulations of length $\tau$ for each anchor point $A^\ell$, yielding $S$ samples $A^{\ell, 1}, \dots, A^{\ell, S} \in \A$ from the distribution $p^\tau_{A^\ell}$.
We employ the maximum mean discrepancy (MMD) to estimate pairwise distances $\Delta_{\ell_1, \ell_2}$ between $p^\tau_{A^{\ell_1}}$ and $p^\tau_{A^{\ell_2}}$, i.e.,
\begin{align}
\label{eq:MMD}
	\Delta_{\ell_1, \ell_2} & := M_{\ell_1, \ell_1} + M_{\ell_2, \ell_2} - 2\ M_{\ell_1, \ell_2},          \\
	M _{\ell_1, \ell_2}     & := \frac{1}{S^2} \sum_{s_1, s_2 = 1}^S \kappa \big(A^{\ell_1, s_1}, A^{\ell_2, s_2} \big),
\end{align}
where $\kappa$ is a positive definite symmetric kernel function~\cite{bittracher_dimensionality_2020}.
The MMD measures the difference between distributions after mapping them into the reproducing kernel Hilbert space induced by $\kappa$, and \Cref{eq:MMD} provides an empirical estimate based on the sampled data.
Here, we use an exponential kernel based on the Hamming distance~$d_H$ between the adjacency matrices, i.e.,
\begin{align}
	\kappa(A,B) & := \exp\!\left(- \frac{d_H(A, B)}{\varepsilon}\right), \label{eq:kernel} \\
	d_H(A, B)   & := \frac{1}{2}\sum_{i,j} \lvert A_{i,j} - B_{i,j} \rvert, \label{eq:d_H}
\end{align}
where $\varepsilon>0$ controls the kernel bandwidth.

Finally, we apply a distance-based manifold learning algorithm to the distance matrix $\Delta \in \R^{L\times L}$.
We choose the diffusion maps method for this purpose \cite{coifman_diffusion_2006}.
This yields an approximation of a low-dimensional embedding of $\M_\tau$ in the form of the coordinates $\varphi(A^1), \dots, \varphi(A^L) \in \R^d$.
The dimension $d$ of the embedding can be inferred from the manifold learning algorithm.
Thus, we obtain the values of the CV $\varphi$ at each anchor point.

\begin{rem}
	The Hamming distance used above is a sensible choice under the assumption that all opinions are distinct, i.e., $\theta_i \neq \theta_j$, since then each node is uniquely identified by its opinion.
    In the present setting, i.e., sampling opinions from continuous distributions like the uniform or normal distribution, this assumption is not restrictive as the probability for coinciding opinions is zero.
	If, however, multiple nodes would share an identical opinion, one may argue that graph isomorphisms (i.e., relabelings of nodes) should be taken into account in the definition of the distance and kernel. While more sophisticated graph kernels that account for such symmetries exist \cite{kriege_survey_2020}, they are typically designed for much smaller graphs.
	Although recent work has reduced the computational cost of common graph kernels \cite{faster_kernels}, they remain computationally impractical for the systems considered here.
\end{rem}

Next, we seek an interpretation of the learned coordinates $\varphi(A^\ell)$ to understand what information the CV contains, and to enable extrapolation to graphs that are not included among the anchor points. Although algorithmic approaches exist for identifying analytical representations of $\varphi(A)$, e.g., by fitting a parametrized candidate function, we found empirically that for the models studied here the consensus measure introduced in the following section provides an excellent fit to the data.

\section{\label{sec:Results} Consensus as a Collective Variable}

We apply the transition manifold approach to generate embeddings for both the ergodic model and the threshold model.
We test four different sets of opinions $(\theta_1,\dots,\theta_N)$: deterministic equidistant, uniformly distributed, normally distributed with mean $\mu=0.5$ and variance $\sigma^2 = 0.2$, as well as a bimodal distribution generated by the mixture of two normal distributions with means $\mu_1 = 0.25$, $\mu_2 = 0.75$ and variance $\sigma^2 = 0.1$.
For the normal and bimodal distributions, samples are truncated to the interval $[0,1]$.

The choice of the bandwidth $\varepsilon$ for the kernel in~\Cref{eq:kernel} has a strong effect on the results and must be considered carefully.
To tune the bandwidth, we use graphs generated by a collection of Watts--Strogatz (WS) models \cite{watts_collective_1998} with different rewiring parameters.
These graphs serve as a proxy for the graphs produced by the dynamics.
Thus, they can be used to generate distances $d_H$ that are typical for the system and, therefore, to determine a suitable bandwidth $\varepsilon$; see \Cref{sec:Bandwith Tuning} for details.

In order to compute the transition manifold, we sampled $L=2000$ anchor points, and for each anchor point we generated $S=500$ burst simulations to estimate the pairwise MMD distances. Provided that $L$ is large enough to cover the relevant state space and $S$ is large enough to accurately sample the transition probability distributions, their specific values have little effect on the resulting embedding.

\begin{figure}
	\centering
	\includegraphics[width=\linewidth]{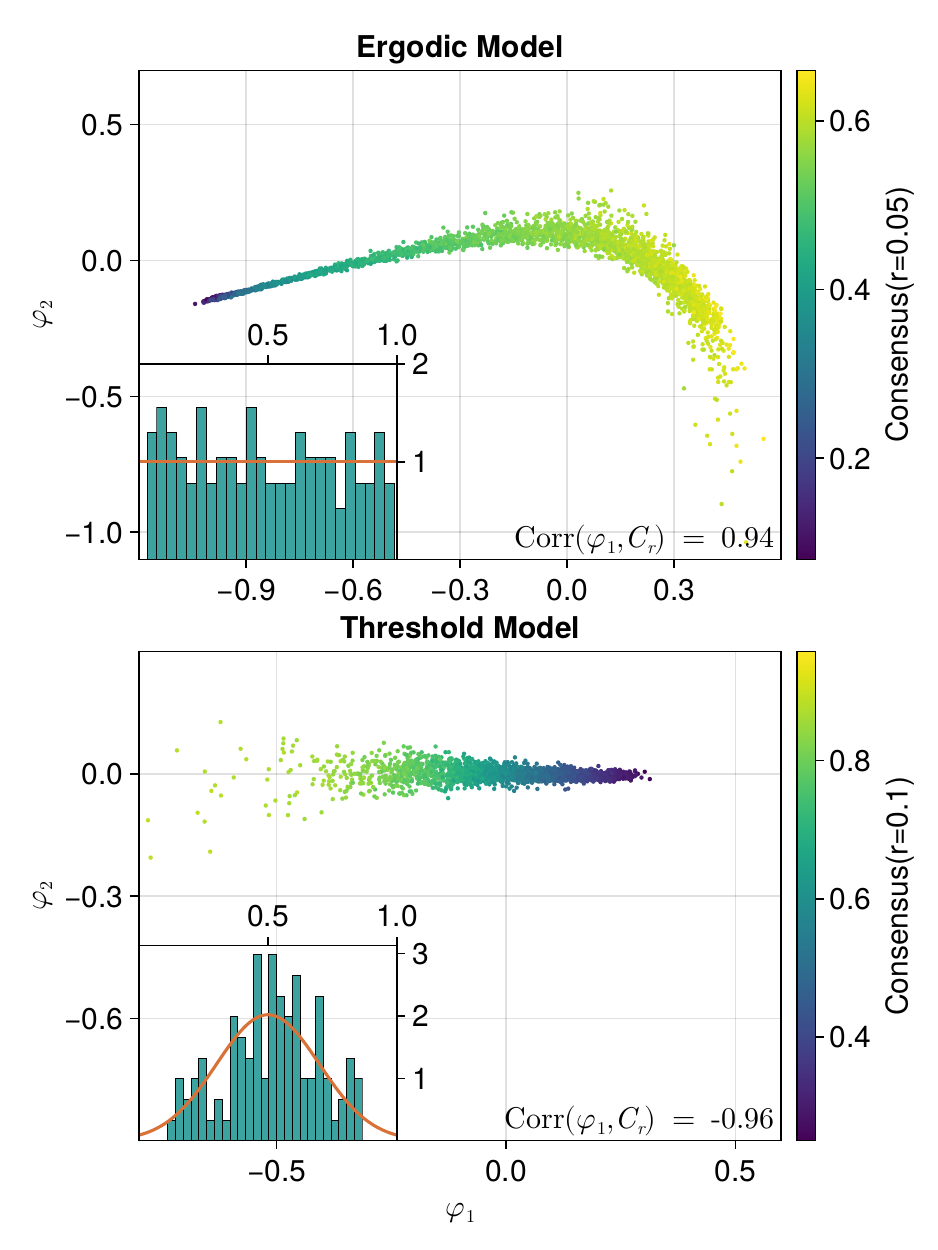}
	\caption{The embedding of the transition manifold for the ergodic model (top) with $p=0.01$ and uniformly distributed opinions (inset), and for the threshold model (bottom) with normally distributed opinions (inset). In both cases, the consensus measure~$C_r$ provides an accurate parametrization of the dominant embedding coordinate,  
        see the high correlation of the color with the dominant coordinate~$\varphi_1$.
		For the threshold model we use the model parameter $r=0.1$, whereas for the ergodic model we find $r=0.05$ to be optimal.
		Parameters: $N=100$, $K=500$, $\lambda=N$, $\varepsilon = K$, $L=2000$, $S=500$,  $T=150$, $\tau=10$, for the ergodic model and $T=10$ and $\tau=0.5$ for the threshold model.}
	\label{fig:smooth_model_cv}
\end{figure}

We found that for both models and all the different opinion distributions, the embedding reveals a one-dimensional transition manifold; see \Cref{fig:smooth_model_cv} for an example and \Cref{sec:Other Opinion Distributions} for additional plots with different model parameters and opinion distributions. The fact that the transition manifold is one-dimensional points to the existence of a one-dimensional model reduction.
Moreover, in~\Cref{fig:smooth_model_cv} we observe that in all cases the dominant diffusion coordinate exhibits a strong correlation with the \textit{consensus measure} $C_r$,
\begin{equation}\label{eq:Consensus}
	C_r(A) := \frac{1}{K}\sum_{i<j} A_{ij}\mathbb{1}_{[0, r]}( \lvert \theta_i -  \theta_j \rvert ),
\end{equation}
indicating that the transition manifold is effectively parametrized by $C_r$. 
The quantity $C_r$ measures the level of consensus via the fraction of concordant edges, i.e., the fraction of edges for which the opinions of the endpoints are closer than the threshold~$r$.
Since the ergodic model does not contain an intrinsic threshold parameter, the optimal value for $r$ is determined numerically by maximizing the correlation between $C_r$ and the dominant coordinate, which yields $r\approx 0.05$ for all examined model parameters, see \Cref{sec:choice_of_threshold_parameter} for details.
In the threshold model, however, the intrinsic opinion threshold also serves as the threshold in~$C_r$.

Note that an almost linear relationship between embedding coordinate and CV, which is implied by the aforementioned high correlation, is not strictly necessary for the CV to be considered good.
A highly non-linear parametrization of the embedding is, in theory, also a suitable CV, but has several practical disadvantages such as sensitivity and numerical instability. 
Thus, we employ the correlation as a metric for the CV.

We have also investigated other commonly used measures of consensus and homophily, e.g., the attribute assortativity \cite{kan_adaptive_2023} which measures the correlation of opinions along edges, see \Cref{sec:Other Opinion Distributions} for details.
We found that the assortativity performs similarly well for the threshold model, but for the ergodic model it has a significantly smaller correlation with the dominant embedding coordinate than the consensus measure $C_r$.
This highlights that collective variables are not unique and that the same collective variable might not be well-suited to capture the dynamics of two models, even if the underlying drivers of the dynamics are similar.
In all the cases that we have investigated, the proposed consensus measure $C_r$ performed well, promising a broader suitability for this class of systems.

\section{\label{sec:MacroModels}Macroscopic Dynamics}
In the following, we highlight two applications of the collective variables once they have been obtained with the approach described above.
First, we demonstrate that further data-driven approaches can be used to identify evolution equations governing the learned collective variables.
Together with the transition manifold approach, this yields a complete data-driven model reduction pipeline that algorithmically constructs an optimal reduced model based on simulation data.
Second, we show that the approach presented in this paper can also serve as a tool for further analytical work.
Since the identified collective variables are designed such that a low-dimensional representation of the system exists, they provide a natural starting point to find closed-form analytical expressions for a reduced model.

\subsection{\label{sec:MacroModelsErgodic}Ergodic Model}
In this section, we apply the \textit{sparse identification of nonlinear dynamics} (SINDy) \cite{Brunton2016} approach to obtain a macroscopic model for the ergodic rewiring model introduced in~\Cref{sec:ergodic_model}.
It was shown in the previous section that the consensus measure $C_r$ is a good CV for this model.
This implies the existence of an approximately Markovian reduced model in the CV coordinates that is largely independent of the microscopic representation and closely approximates the projected dynamics.
This reduced model is, in general, stochastic.
However, in many applications the expected value $\E[\rv{C}_r(t)]$ is of particular interest, where $\rv{C}_r(t) := C_r(\rv{A}(t))$ denotes the projected process.
Since this quantity is deterministic, it is natural to search for an ODE to describe its time evolution.
Because $C_r$ is a good CV, it is ensured that two trajectories of the projected process starting from the same initial value $\rv{C}_r(0)$ exhibit very similar macroscopic evolution, even though the associated microscopic initial states $\rv{A}(0)$ may differ substantially.
If $C_r$ were not a good CV, the problem of finding an ODE describing $\E[\rv{C}_r(t)]$ would be ill-posed since identical initial values could lead to different projected trajectories.
It is, however, not generally guaranteed that the macroscopic dynamics can be represented by an ODE, even if the CV is good.
We now employ the SINDy method, which uses regularized regression to construct an ODE that best fits the simulation data from a dictionary of basis functions, and we will see that, in this case, an ODE is indeed well-suited for defining the macroscopic system.

\begin{figure}
	\centering
	\includegraphics[width=1\linewidth]{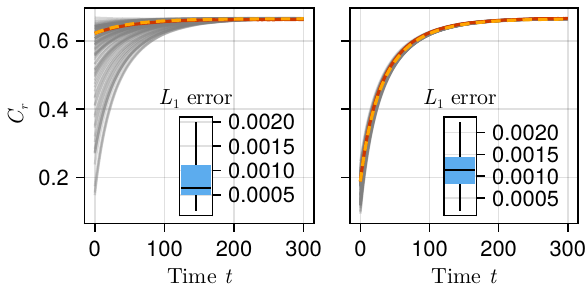}
	\caption{
        SINDy results for the training data (left) and validation data (right), using the CV $C_r$ with $r = 0.05$.
		Each plot shows the 100 trajectories in the data set (gray).
        For each trajectory we have computed the error of the learned ODE \cref{eq:SINDy_ODE} in time-normalized $L_1$-norm \cref{eq:L1-error}; the error statistics are visualized in the boxplot. 
        We show the trajectory with largest error (red) and the associated ODE solution (orange dashed).
    }
	\label{fig:sindy_ergodic}
\end{figure}

We examine the ergodic model with acceptance probability $p=0.01$ and uniformly distributed opinions $\theta_i \sim \mathcal{U}([0,1])$.
First, we sample 100 initial graphs using the same sampling strategy as described earlier, i.e., by first sampling a random graph and then simulating the dynamics for a randomly chosen time.
Then, for each of the resulting 100 graphs, we conduct 1000 simulations of length $T=300$ to approximate $\E[\rv{C}_r(t)]$ for $t\in[0,T]$ and $r=0.05$, which was the optimal value for~$r$ determined in the previous section.
Hence, we obtain 100 training trajectories estimating $\E[\rv{C}_r(t)]$ for different initial conditions.
Finally, we apply SINDy with monomials of degree less than or equal to two as the dictionary functions.
This results in the ODE
\begin{equation} \label{eq:SINDy_ODE}
	\frac{\dif}{\dif t} \bar{C}_r(t) = 0.027 - 0.062\ \bar{C}_r(t) + 0.033\ \bar{C}_r(t)^2.
\end{equation}
We quantify its accuracy using a time-normalized $L_1$ error, i.e., if $\bar{C}_r(t)$ denotes the ODE solution of \Cref{eq:SINDy_ODE} and $\E[\rv{C}_r(t)]$ the trajectory from the data set, both started in the same initial state, we measure the error
\begin{equation} \label{eq:L1-error}
	L_1(T) := \frac{1}{T} \int_0^T \big\lvert \bar{C}_r(t) - \E[\rv{C}_r(t)] \big\rvert\, \dif t.
\end{equation}
\Cref{fig:sindy_ergodic} (left) illustrates that the learned ODE given in~\Cref{eq:SINDy_ODE} matches the training trajectories with very high accuracy.
Even the largest observed $L_1$-error is only approximately $0.002$ per time unit.

We further verify that the learned ODE generalizes well to initial conditions that were not part of the training data set by constructing initial graphs using the stochastic block model, see \Cref{fig:sindy_ergodic} (right).
For the stochastic block model, we define three equally sized blocks and vary the intra- and inter-block edge densities to generate different network structures.
This demonstrates that the CV and the learned ODE provide an accurate approximation not only for the initial conditions from the training set, but also for other graphs that are unlikely to occur under the dynamics.

We also applied this method to the other opinion distributions considered earlier (normal and bimodal).
This resulted in structurally similar ODEs, but with different coefficients.
Furthermore, the approximation error for both the training and the validation trajectories was comparable to that observed in the presented case.

\subsection{\label{sec:Graphon}Threshold Model}
In this section, we derive a closed-form ODE for the evolution of the graphon approximation of $C_r(t)$ for the threshold model, demonstrating that the data-driven approach correctly identified the relevant collective variable. To analyze the graph dynamics in the continuum limit, we formulate a balance equation for the \textit{graphon} 
\begin{equation}
    g_t: [0,1]^2 \to [0,1],
\end{equation}
where $g_t(x,y)$ denotes the local edge density between the continuum node labels $x$ and $y$ at time $t$. 
Since the underlying graphs are undirected, the graphon is symmetric
\begin{equation}
    g_t(x,y)=g_t(y,x).
\end{equation}
We present a concise derivation of the graphon approximation here; an extensive derivation containing all technical details can be found in~\Cref{sec:Graphon CV}.

The rewiring processes studied here induce a redistribution of edge density across the graph while conserving the total number of edges. Thus, the evolution of the graphon can naturally be formulated through a gain--loss balance equation balancing edge formation and dissolution. In particular, the total edge density
\begin{equation}
    e := \int_{[0,1]^2} g_t(x,y)\, \dif x\,\dif y = \frac{K}{\binom{N}{2}},
\end{equation}
remains constant over time.
Let the agent opinions be given by a cumulative distribution function $F: [0,1] \to [0,1]$; i.e., $F(\theta_0)$ is the fraction of agents having opinion~$\theta \le \theta_0$. Recalling that agents are ordered according to their opinions, we can map the continuous node label $x \in [0,1]$ to an opinion $\theta(x)$ via the quantile function
\begin{equation}
	\theta(x) := \inf\{\theta \in [0,1] : F(\theta) \ge x\}.
\end{equation}
We define the discordant domain $\mathcal{D} \subset [0,1]^2$ as the set of all node pairs whose opinion difference exceeds the given homophily threshold~$r$, i.e.,  $\mathcal{D} := \{ (x, y) \in [0,1]^2 \mid |\theta(x) - \theta(y)| > r \}$.
The complement of the discordant domain is the concordant domain $\mathcal{C} := \mathcal{D}^c \subset [0,1]^2$, which contains all concordant node pairs.
Integrating over $\mathcal{C}$ yields the total density of concordant edges
\begin{equation}
  c(g_t):= \frac{1}{e}\int \mathbb{1}_{\mathcal{C}}(x,y)g_t(x, y)\,\dif x\,\dif y,
\end{equation}
which is the continuum version of the consensus measure $C_r$, cf. \Cref{eq:Consensus}. 

In the graphon limit the rewiring probability in~\Cref{eq:rewiring_probability} takes the form
\begin{multline} \label{eq:rewiring_probability_limit}
    p(x, y; g_t) := \\
    \frac{1}{Z(x; g_t)} (1-|\theta(x) - \theta(y)|)(1-g_t(x,y)),
\end{multline}
where 
\begin{equation}
    Z(x; g_t) := \int_0^1 (1-|\theta(x) - \theta(y)|)(1-g_t(x,y)) \, \dif y.
\end{equation}
In the microscopic model with $N$ nodes, let the rate of rewiring events be given by $\lambda_N$ (denoted $\lambda$ in previous sections). A single rewiring event alters exactly one edge, which corresponds to a change in the overall edge density of magnitude $1/\binom{N}{2}$.  As we pass to the graphon limit, the magnitude of individual jumps vanishes; thus, for 
a non-degenerate limit, the rate has to be rescaled to $\lambda_N = \lambda \binom{N}{2} $ to ensure that the density change remains finite and non-zero.

This scaling of the interaction rate results in the macroscopic description in an effective rate $\lambda$ that is
constant with respect to the system size $N$ and allows us to formulate the \textit{balance equation}
\begin{align}\label{eq:balance_main}
    \frac{\partial g_t}{\partial t}(x,y) & = \Big( \tfrac12 [ p(x,y;g_t) d(x; g_t) + p(y, x;g_t) d(y; g_t)] \nonumber \\
	                                     &\phantom{=} \quad\  - \mathbb{1}_{\mathcal{D}}(x,y) g_t(x,y)\Big) \frac{\lambda}{e\, (1-c(g_t))},
\end{align}
where 
\begin{equation}
    d(x; g_t)            := \int_0^1 \mathbb{1}_\mathcal{D}(x, y) g_t(x, y) \, \dif y
\end{equation}
is the \textit{discordant degree function}.
Each summand in the gain term (first line of~\eqref{eq:balance_main}) reflects the fact that for an edge $(x,y)$ to appear, one of its ends needs to lose an old neighbor and rewire to the new one.
The indicator function in the loss term (second line) reflects that only discordant edges can be rewired.

In order to obtain an approximation for the dynamics of the consensus measure, we need to capture the non-linear ``crowding'' effect that slows down the growth of $c_t := c(g_t)$ as the concordant domain $\mathcal{C}$ saturates, making rewiring into $\mathcal{C}$ less likely.
We approximate $g_t$ as a step function based on the average local density within the two domains
\begin{equation}
	g_t(x,y) \approx \begin{cases}
	  \rho_{\text{dis}}(c_t) := \frac{e(1-c_t)}{| \mathcal{D}|}       & \text{if } (x,y) \in \mathcal{D}   \\
	  \rho_{\text{con}}(c_t) := \frac{ec_t}{| \mathcal{C} |} & \text{if } (x,y) \in \mathcal{C}
	\end{cases}
\end{equation}
where $| \mathcal{C} |, |\mathcal{D}| \in [0,1]$ are the geometric areas of the discordant and concordant domains, respectively, such that $| \mathcal{C} | + |\mathcal{D}| = 1 $.

If $c_t$ is a good collective variable, then all graphons with the same consensus measure will behave similarly, motivating the choice of the above representative most suited for further
analysis.
With this ansatz and under the assumptions that the graphs belong to a sequence that converges to a graphon limit, the consensus measure follows the ODE
\begin{equation}\label{eq:graphon_ODE}
  \frac{\dif c_t}{\dif t}  = \frac{\lambda}{e} \left[ 1 - \mathcal{S}(\alpha(c_t)) \right].
\end{equation}
Here, the quantity $\smash{ \alpha(c) := \frac{1-\rho_{\text{con}}(c)}{1-\rho_{\text{dis}}(c)} }$ captures the imbalance of the saturations in the concordant and discordant domains, and $\mathcal{S}(\alpha)$ is a \textit{stagnation factor} which reduces the growth rate of $c_t$. More precisely,
\begin{equation}
  \mathcal{S}(\alpha) := \frac{1}{|\mathcal{D}|} \int_0^1 \frac{\ell_\mathcal{D}(x)}{1+ R(x)\,\alpha} \, \dif x,
\end{equation}
where $\ell_\mathcal{D}(x) := \int_0^1 \mathbb{1}_\mathcal{D}(x,y) \,\dif y$ describes the number of potential discordant neighbors, and
\begin{equation}
  R(x) := \frac{\int_0^1 \mathbb{1}_{\mathcal{C}}(x,y)(1-\|\theta(x)-\theta(y)\|) \,\dif y}{\int_0^1 \mathbb{1}_\mathcal{D}(x,y)(1-\|\theta(x)-\theta(y)\|) \,\dif y}
\end{equation}
captures the ratio of the general affinity of node $x$ to form a concordant edge versus a discordant edge.
If the opinions are distributed according to a uniform distribution, the stagnation factor $\mathcal{S}(\alpha)$ has an algebraic solution.
In general, however, this integral has to be evaluated numerically during integration.

\begin{figure}
	\centering
	\includegraphics[width=\linewidth]{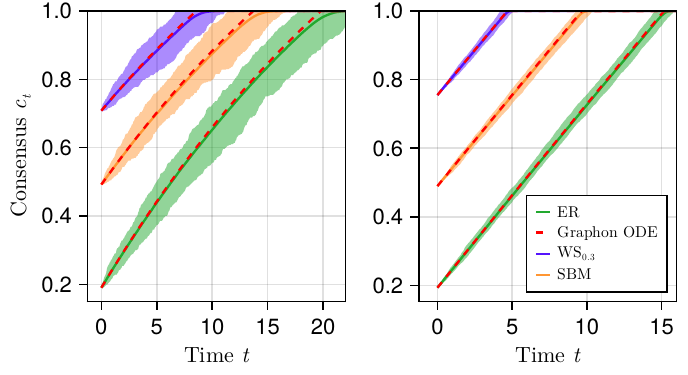}
	\caption{
	  Evolution of the consensus measure $c_t$ given by \cref{eq:graphon_ODE} compared to $\E[\rv{C}_r(t)]$ in the sparse regime, $e\leq|\mathcal{C}|$, for different initial graphs: Erdős-Rényi (ER), Watts--Strogatz (WS; rewiring parameter $0.3$), and a three-block stochastic block model (SBM) with uniformly distributed opinions.  Dynamics for denser networks with $N=100$ nodes and $K=500$ edges (left). Dynamics for moderately sparse networks with $N=1000$ nodes and $K=4995$ edges (right). Dashed lines represent the continuum limit solution, solid lines indicate the mean of 300 independent realizations of the threshold model, and shaded envelopes show the extrema. The normalized $L_1$ errors are less or equal than  $6 \times 10^{-3}$ (left) and  $10^{-3}$ (right). Parameters: $\lambda_N = N$ and $r=0.1$.
	}
	\label{fig:graphon}
\end{figure}

In \Cref{fig:graphon}, we compare solutions of \Cref{eq:graphon_ODE}
to realizations of the threshold model for three different initial network topologies: Erdős--Rényi, Watts--Strogatz, and a stochastic block model.
We find that the graphon ODE approximates $\E[\rv{C}_r(t)]$ very well for large graphs with $N=1000$ nodes  and $K=4995$ edges,
see the right panel of \Cref{fig:graphon}.
Even for smaller networks with $N=100$ and $K=500$ edges, which are the same parameters that were used in~\Cref{sec:Results} to identify the consensus measure as a collective variable, and typically do not fall into the relevant range for mean-field approximations, the derived ODE captures $\E[\rv{C}_r(t)]$ reasonably well, see the left panel of \Cref{fig:graphon}. The ODE also gives a good approximation for normally distributed opinions, see \Cref{fig:graphon_appendix} in~\Cref{sec:Graphon CV}.

\begin{figure}
	\centering
	\includegraphics[width=\linewidth]{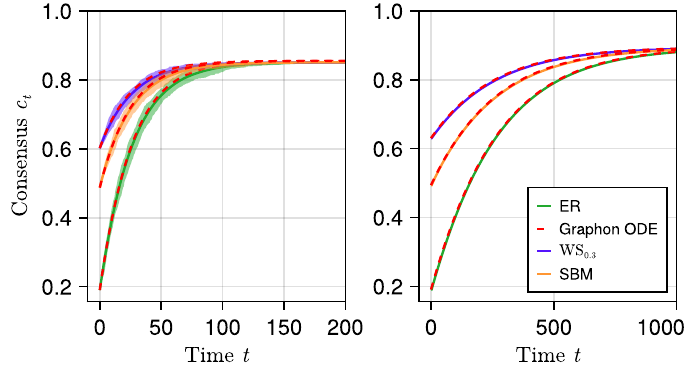}
	\caption{Evolution of the consensus measure $c_t$ given by \cref{eq:graphon_ODE} compared to $\E[\rv{C}_r(t)]$ in the dense regime, $e > |\mathcal{C}|$. All parameters as in~\cref{fig:graphon} except $K = 1100$ (left) and $K =
	106000$~(right) and 100 realizations were used. The normalized $L_1$ errors are less or equal than  $5 \times 10^{-3}$ (left) and  $2\times 10^{-3}$~(right).}
	\label{fig:graphon_dense_regime}
\end{figure}

\Cref{fig:graphon} shows the sparse regime of the threshold model, $\binom{N}{2}e= K \leq K_c = \binom{N}{2}|\mathcal{C}|$, where there exists an absorbing state for which $c_t = 1$, cf. \Cref{sec:threshold_model}.
The derived ODE in \Cref{eq:graphon_ODE} is also valid in the opposite dense regime, $e > |\mathcal{C}|$, where there are no absorbing states, i.e., the consensus measure $c_t$ saturates at a level smaller than 1, which is shown in \Cref{fig:graphon_dense_regime}.

The fact that the ODE captures the dynamics of the consensus measure $\rv{C}_r(t)$ in both the sparse and the dense regime, and for different opinion distributions and different initial network topologies, confirms that $C_r$ is indeed a good collective variable.

\section{\label{sec:Conclusion} Conclusion}
In this work, we identified low-dimensional collective variables for systems of time-evolving networks, focusing on two representative network rewiring processes driven by homophily with respect to static node attributes, which we called opinions.
The collective variables were obtained by the data-driven transition manifold approach.
Across a broad range of opinion distributions, our method consistently found a one-dimensional collective variable that captures the essential dynamics of the system.
We found that this collective variable is accurately parametrized by a consensus measure that quantifies the fraction of concordant edges, i.e., edges connecting nodes whose opinions differ by less than a prescribed threshold.
 Note that the opinion distribution can be viewed as a high-dimensional, and in the large-network limit even infinite-dimensional, parameter of the model. The observed robustness of the transition manifold and associated collective variable with respect to these parameters suggests that our findings extend beyond the specific examples considered here.
Moreover, the interpretability of the collective variable allows for considering it as a candidate observable for similar processes, and its analytic form opens the way for theoretic analysis and reduced modelling.

Beyond identification of the collective variable, we showed how it can be used to construct reduced macroscopic models. For the ergodic rewiring model, we combined the transition manifold approach with sparse regression and obtained a simple ODE that accurately reproduces the evolution of the consensus measure. For the threshold model, we derived a graphon-based macroscopic description and obtained a closed-form evolution equation for the consensus measure. The excellent agreement between the reduced models and the projected microscopic dynamics, across different network topologies and opinion distributions, provides independent validation that the identified consensus measure is indeed a suitable collective variable.

Classical approaches to model reduction for adaptive networks, such as moment closure schemes \cite{demirel_moment-closure_2014, song_evolutionary_2022,kiss_mathematics_2017} and mean-field theories \cite{Gkogkas2022,berner_adaptive_2023}, provide valuable insight into macroscopic network dynamics. However, these methods typically require strict modelling assumptions or analytical choices for the system under consideration. The transition manifold approach complements such methods by providing a largely data-driven route to identifying suitable macroscopic coordinates directly from simulation data. In particular, a key advantage of the approach is that physically meaningful collective variables can emerge from the dynamics rather than being prescribed a priori.

Several directions for future work appear promising. The most natural extension is the application of the framework to fully adaptive network models in which node states and network structure co-evolve.
More generally, transition manifold learning may provide a systematic route towards analytical model reductions of adaptive network dynamics in a wide range of applications.

\subsection*{Acknowledgement}
This work has been partially funded by the Deutsche Forschungsgemeinschaft (DFG, German Research Foundation) through Project No.~546032594 and under Germany's Excellence Strategy – The Berlin Mathematics Research Center MATH+ (EXC-2046/1, EXC-2046/2, project ID: 390685689).

\clearpage
\appendix
\crefalias{section}{appendix}
\clearpage
\section{\label{sec:Bandwith Tuning} Bandwidth Tuning}
The choice of bandwidth parameter~$\varepsilon$ in \eqref{eq:kernel} plays a major role for a successful embedding of the transition manifold.
If the chosen bandwidth is too large, the kernel takes values close to $1$ and hence the MMD is always close to zero.
This results in an unsuccessful embedding, since all the points collapse to a single point or a small ball.
If, on the other hand, the chosen bandwidth is too small, the kernel becomes too localized and the MMD of any two (different) distributions becomes large.

In order to tune the bandwidth we use the one-parameter family of graph distributions generated generated by the Watts--Strogatz model with rewiring probability $p$, which we denote by WS$_p$.
This family provides a convenient proxy for the graphs produced by the processes considered here, since it covers uniform Erd{\H o}s--R\'enyi networks for $p\approx 1$ and ring lattices for $p\approx 0$.
The ring lattices are similar to the graphs generated by the invariant measure of the ergodic models and the absorbing states of the threshold model. In contrast, the ER networks resemble the graphs generated by the process when the intermediate integration time $t$ used in the anchor sampling is close to zero.

To tune the bandwidth efficiently, we derive an analytical approximation of the MMD between Watts--Strogatz graphs. This avoids repeatedly estimating MMD values from Monte Carlo samples, which would be computationally expensive. 
First, we partition the set of possible edges into those belonging to the regular ring lattice and those belonging to the rewiring zone (outside of the ring lattice).
For a Watts--Strogatz graph with $N$ nodes and mean degree $k$, the number of possible edges in the ring lattice is $K := \frac{N k}{2}$, and the possible number of edges in the rewiring zone is $\smash{ \tilde{K} := \binom{N}{2} - K }$.
Subsequent rewirings are sampled without replacement, but for simplicity, we approximate them as independent Bernoulli trials. This approximation is justified because the probability of selecting a node that is already connected to the current node is 
roughly $p\frac{d_i-1}{N-1}$, where $d_i$ is the current degree of the node. Since $d_i \approx k$, this probability is of order $\smash{ \mathcal{O}(\frac{k}{N}) }$, which is negligible if $N \gg k$. Therefore, we treat the existence of edges as independent Bernoulli trials with probability $q := 1-p$ in the ring graph and $\tilde{q} := \frac{pK}{\tilde{K}}$ in the rewiring zone.
Each possible edge contributes to the Hamming distance $d_H$ if it exists in one graph and not in the other.
Hence, the probability that any possible node pair contributes is given by $\phi(x,y) := x(1-y) + y(1-x)$, where $x,y$ are the probabilities of the specific edge existing in each of the graphs.
Let $\rv{H} := d_H(\rv{A}^{(1)},\rv{A}^{(2)})$ denote the Hamming distance between two independently sampled Watts--Strogatz graphs with rewiring probabilities $p_1$ and~$p_2$. Since we treat the edges as independent Bernoulli trials, the mean $\mu$ and variance $\nu^2$ of the Hamming distance $\rv{H}$ are approximately given by
\begin{align}
	\mu   & \approx K \phi(1-p_1, 1-p_2) + \tilde{K}\phi \left( \frac{p_1 K}{\tilde{K}}, \frac{p_2 K}{\tilde{K}}\right)                                                            \\
	\nu^2 & \approx K\phi(1-p_1, 1-p_2)(1-\phi(1-p_1, 1-p_2)) \nonumber                                                                                                                      \\
	      & + \tilde{K} \phi\left(\frac{p_1 K}{\tilde{K}}, \frac{p_2 K}{\tilde{K}} \right)\left(1-\phi\left(\frac{p_1 K}{\tilde{K}}, \frac{p_2 K}{\tilde{K}} \right)\right).
\end{align}
With $\binom{N}{2}$ possible edges taken as individual Bernoulli trials, we can exploit the central limit theorem to approximate the distribution of the Hamming distance $\rv{H}$ between the two random WS graphs by the normal distribution, $\rv{H} \sim \mathcal{N}(\mu, \nu^2)$.
Finally, to compute the expectation of the kernel evaluations, we make use of the analytic form of the moment generating function:
\begin{equation}\label{eq:ws_kernel_exp}
	\E\left[ \exp\left( -\frac{\rv{H}}{\varepsilon}\right)\right] = \exp \left( -\frac{\mu}{\varepsilon} + \frac{\nu^2}{2\varepsilon^2}\right).
\end{equation}
We validated this approximation against a numerical estimate computed using \Cref{eq:MMD}.
The resulting relative error was of order $10^{-2}$, which is sufficiently small for bandwidth selection. With the analytically tractable proxy, we can tune the bandwidth in a computationally efficient manner.

\begin{figure}
	\centering
	\includegraphics[width=\linewidth]{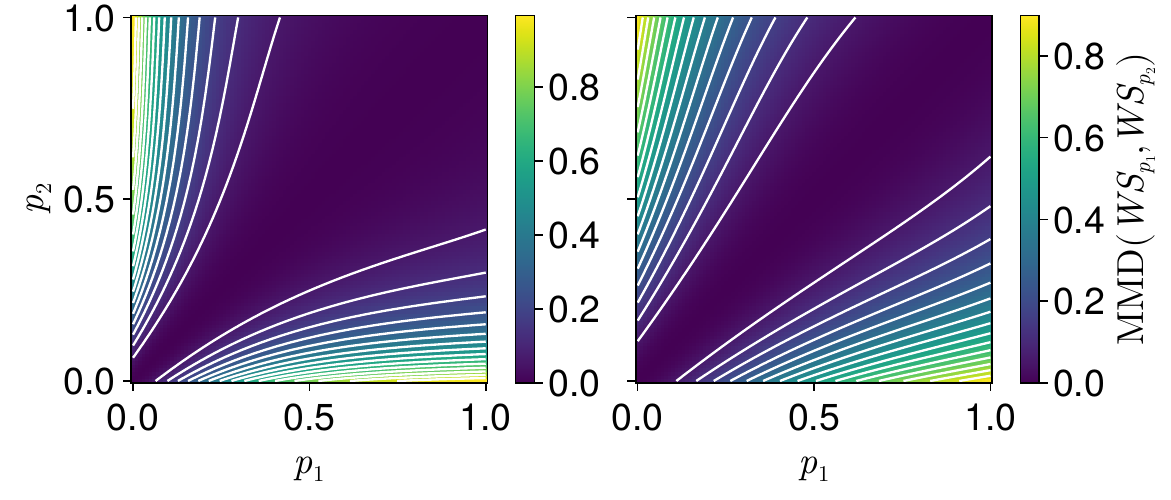}
	\caption{Comparison of the MMD for two WS models with different rewiring parameters $p_1$ and $p_2$. If the bandwidth is too small ($\varepsilon = \tfrac K2$), the MMD cannot distinguish between relevant distributions (left). A good bandwidth ($\varepsilon = K$) resolves the space while keeping large range of MMD values (right).}
	\label{fig:tuning}
\end{figure}

We want to select a bandwidth $\varepsilon$ that gives us maximal distinguishability across the entire range of graphs. 
The modelling assumption for this purpose is that the graphs relevant for the process (and hence for the transition manifold computation) are well represented by Watts--Strogatz graphs WS$_p$ with rewiring parameter~$p\in[0,1]$.
Thus, we use \Cref{eq:ws_kernel_exp} to calculate the MMD diagrams shown in~\Cref{fig:tuning} for exemplary bandwidths, without running the full MMD estimation from samples.
\Cref{fig:tuning} compares the pairwise MMD values obtained for an overly small bandwidth (left) and a well-tuned bandwidth (right). The MMD is evaluated for pairs of Watts--Strogatz models with rewiring parameters $(p_1, p_2)$.
If the bandwidth $\varepsilon$ is too small, there is a large collection of $(p_1, p_2)$ pairs that cannot be resolved properly. In this case, the MMD cannot distinguish between, for example, a WS$_{0.6}$ model and a WS$_{1}$ model, while we get extremely good resolution when one of the rewiring parameters is close to~$0$. In order to obtain an informative transition manifold embedding, the MMD needs to resolve all dynamically relevant states, which is illustrated in the right panel of \Cref{fig:tuning}. Increasing the bandwidth beyond $\varepsilon = K$ results in even more uniform gradients, but also in less magnitude overall, which again results in a lower resolution.

\section{\label{sec:Other Opinion Distributions} Additional Opinion Distributions and Alternative Collective Variables}

\begin{figure}
	\centering
	\includegraphics[width=.95\linewidth]{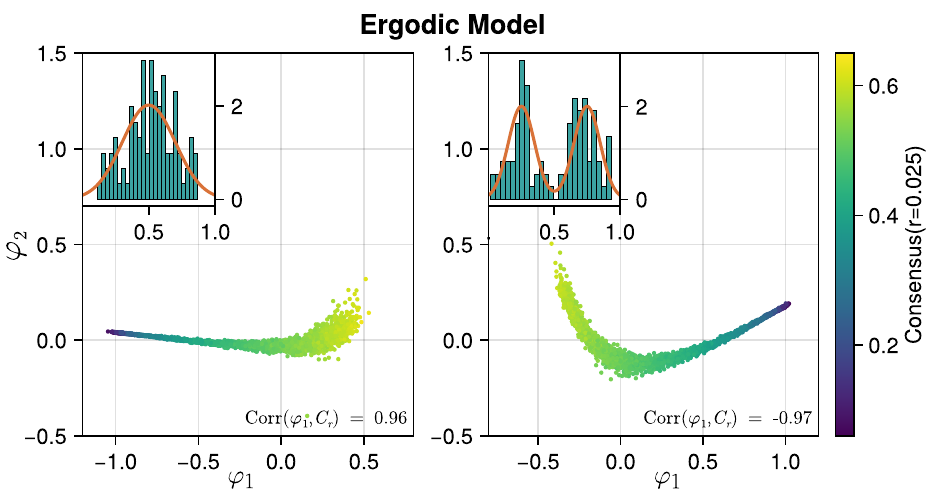}
	\caption{The transition manifold embedding yields the same collective variable for the ergodic model for different opinion distributions. Normal distribution (left) with mean $\mu=0.5$ and variance $\sigma^2 = 0.2$; bimodal distribution (right) consisting of two normal distributions with $\mu_1 = 0.25$, $\mu_2 = 0.75$, and variances $\sigma_1^2=\sigma_2^2=0.1$, see insets. Other parameters as in~\cref{fig:smooth_model_cv}.}
	\label{fig:normal_bimodal_ergotic_cvs}
\end{figure}

\begin{figure}
	\centering
	\includegraphics[width=\linewidth]{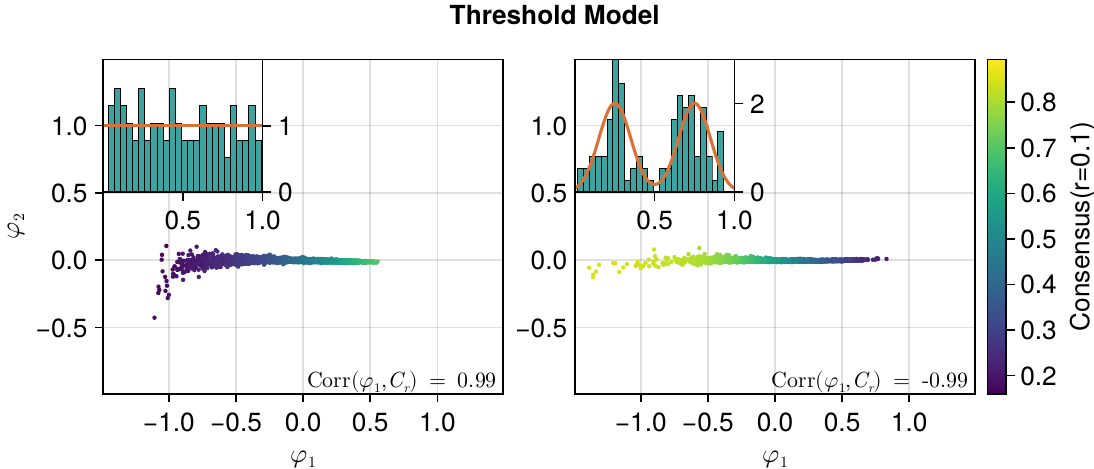}
	\caption{The transition manifold embedding yields the same collective variable for the threshold model for a uniform (left) and bimodal opinion distribution (right) as in~\cref{fig:normal_bimodal_ergotic_cvs}. Other parameters are as in~\cref{fig:smooth_model_cv}.}
	\label{fig:threshold_cvs}
\end{figure}

In the main text, we present the transition manifold embedding for the ergodic model with uniformly distributed opinions and for the threshold model with normally distributed opinions, see \Cref{fig:smooth_model_cv}. In both cases, the embedding identifies the consensus measure $C_r$ as a suitable collective variable. Here, we extend the analysis to additional opinion distributions. In particular, \Cref{fig:normal_bimodal_ergotic_cvs,fig:threshold_cvs} show the embeddings for the ergodic model with normally distributed and bimodally distributed opinions, and for the threshold model with uniformly distributed and bimodally distributed opinions, respectively. 
In all cases, the transition manifold embedding yields an approximately one-dimensional manifold, whose leading diffusion-map coordinate $\varphi_1$ exhibits a strong correlation with the consensus measure. For the ergodic model, we obtain $|\mathrm{Corr}(\varphi_1,C_r)|\geq 0.96$ for both distribution types, see \Cref{fig:normal_bimodal_ergotic_cvs}.
Similarly, \Cref{fig:threshold_cvs} shows that the same behavior persists for the threshold model, where we observe correlations of $|\mathrm{Corr}(\varphi_1, C_r)| = 0.99$ for both opinion distributions.

These results demonstrate that the consensus measure provides a robust parametrization of the transition manifold across a broad range of opinion distributions. This suggests that the transition manifold approach may remain effective in more general adaptive network models, where the opinion distribution evolves together with the network structure.

We have also investigated other candidates for collective variables. A natural choice is the \textit{attribute assortativity}, a commonly used measure of homophily in networks~\cite{kan_adaptive_2023}, defined by
\begin{align}\label{eq:assortativity}
  \frac{\sum_{j, k = 1}^{N}{A_{jk}(\theta_j -\bar{\theta})(\theta_k-\bar{\theta})}}{\sum_{j=1}^N{d_j(\theta_j-\bar{\theta})^2}} \in [-1,1]
,\end{align}
where $\bar{\theta} := \frac{1}{2K}\sum_{j=1}^N d_j \theta_j$ denotes the degree-weighted mean opinion, $d_j$ is the degree of node $j$, and $K$ is the total number of edges. Positive assortativity values indicate that nodes with similar opinions are more likely to be connected, while negative values indicate a preference for connections between nodes with dissimilar opinions. 

\begin{figure}
	\centering
	\includegraphics[width=\linewidth]{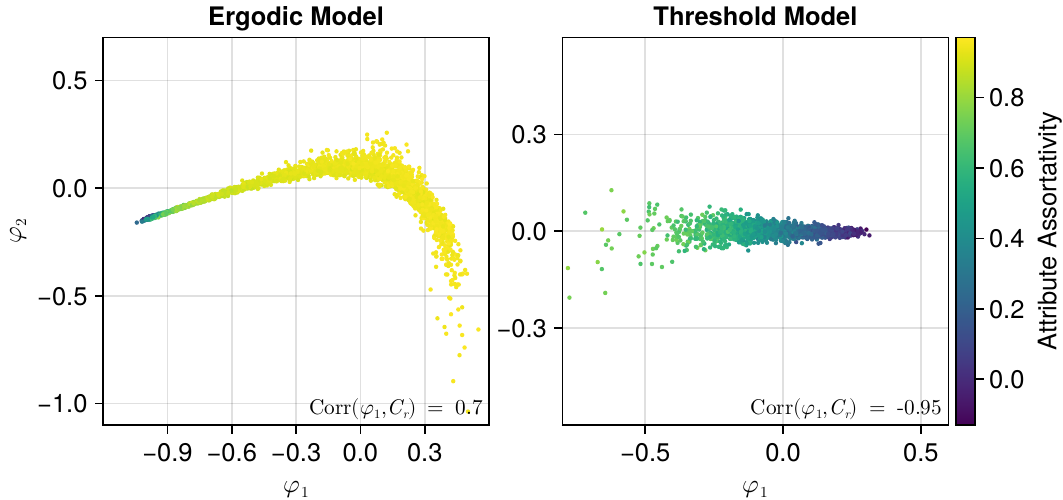}
	\caption{The transition manifold embedding as shown in~\cref{fig:smooth_model_cv}, with the anchor points colored according to their attribute assortativity. For the ergodic model the assortativity is less suitable as a collective variable since it has a much lower correlation with $\varphi_1$ than the consensus measure. For the threshold model, the assortativity is also a good collective variable.}
	\label{fig:other_cvs}
\end{figure}

In \Cref{fig:other_cvs}, we show the same transition manifold embeddings as in~\Cref{fig:smooth_model_cv}, but with the node colors indicating the attribute assortativity. The strong correlation between the assortativity and the dominant embedding coordinate $\varphi_1$ shows that the assortativity provides a good parametrization of the transition manifold for the threshold model.
However, in the ergodic model, the assortativity has a much lower correlation with $\varphi_1$ than the consensus measure $C_r$, cf. \Cref{fig:smooth_model_cv}.
This highlights that even though both models have homophilic interactions as a key dynamical mechanism, the collective variable that best captures the essential dynamics may differ.

\section{\label{sec:choice_of_threshold_parameter} Choice of Threshold Parameter for the Consensus Measure}

\begin{figure}[h]
	\centering
	\includegraphics[width=.75\linewidth]{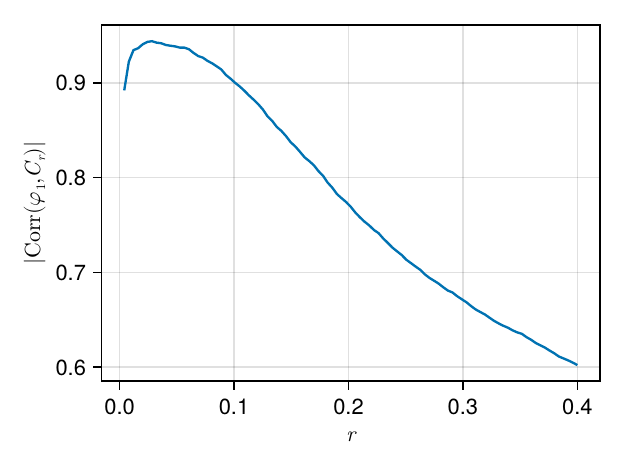}
	\caption{Correlation of $\varphi_1$ with $C_r$ for the transition manifold shown in~\cref{fig:smooth_model_cv}, for different choices of $r$. The correlation exhibits a maximum around $r\approx 0.05$.}
	\label{fig:optimal_r}
\end{figure}

Unlike in the threshold model, where the threshold $r$ is part of the model definition, the ergodic model provides no natural choice of  $r$. We therefore determine $r$ by maximizing the correlation $|\text{Corr}(\varphi_1, C_r)|$ between the consensus measure and the leading diffusion-map coordinate.
\Cref{fig:optimal_r} shows the typical behavior of $| \text{Corr}(\varphi_1, C_r)|$ for different thresholds.
It has a unique maximum around $r\approx 0.05$, which is the value chosen in the main text.

\section{\label{sec:Graphon CV} Graphon CV}
To analyze the graph's evolution in the continuum limit, we formulate a balance equation using \textit{graphons}. This allows us to derive an approximate evolution equation for the consensus measure, which was identified as the relevant collective variable and therefore provides the natural basis for a reduced model. 
While a rigorous proof of convergence from the microscopic threshold model to the balance equation as $N\rightarrow\infty$ is beyond the scope of this work, our numerical results demonstrate that the derived ODE accurately captures the underlying dynamics. Consequently, this model reduction provides a reliable tool for predicting the long-term behavior of the system.

The state of a large graph $A^N_t$ with $N$ nodes at time $t$ is approximated by a graphon $g_t: [0,1]^2 \to [0,1]$, representing the local edge density between the continuum node labels $x$ and $y$.
This graphon limit holds for certain sequences of dense graphs \cite{lovasz_limits_2006, borgs_convergent_2012}.
Since the rewiring processes studied here conserve the total number of edges, the global edge density
\begin{equation}
    e := \int_{[0,1]^2} g_t(x,y)\, \dif x\,\dif y
\end{equation}
remains constant over time. Consequently, the evolution of $g_t$ can naturally be formulated through a gain--loss balance equation balancing edge formation and dissolution across the graph.

First, we define a continuous mapping from node labels $x \in [0,1]$ to opinions. Let the distribution of opinions be described by a cumulative distribution function $F: \mathbb{R} \to [0,1]$ with $F(0)=0$ and $F(1)=1$. We map the continuous node label $x$ to an opinion $\theta(x)$ via the quantile function
\begin{equation}
	\theta(x) := \inf\{\theta \in [0,1] : F(\theta) \ge x\}.
\end{equation}
	Given a homophily threshold $r$, the discordant domain $\mathcal{D} \subset [0,1]^2$ is the set of all node pairs whose opinion difference exceeds this threshold,
\begin{equation}
	\mathcal{D} := \{ (x, y) \in [0,1]^2 \mid |\theta(x) - \theta(y)| > r \}.
\end{equation}
Conversely, the concordant domain $\mathcal{C} := \mathcal{D}^c \subset [0,1]^2$ contains all node pairs whose opinion difference does not exceed the threshold.

Integrating a graphon $g$ over $\mathcal{C}$ gives us the global density of concordant edges
\begin{equation}
  c(g) := \frac{1}{e}\int_{\mathcal{C}} g(x, y)\,\dif x\,\dif y  \in [0,1],
\end{equation}
which is the continuum version of the collective variable~$C_r(A)$,  see \Cref{eq:Consensus}.

Next, we formulate the balance equation for the graphon process. The rewiring dynamics is driven by a rewiring process occurring at a total rate of $\lambda$, obtained from an appropriate rescaling of the microscopic rate $\lambda_N$, as discussed in the main text.
An edge is chosen for rewiring by sampling uniformly from the set of all discordant edges. Hence, the probability density for
an edge at $(x, y)$ to be selected is given by $s(x,y;g)g(x,y)$, where the weight function $s(x,y;g)$ is defined, for $c(g)< 1$, as
\begin{equation}
	s(x, y;g) := \frac{\mathbb{1}_{\mathcal{D}}(x, y)}{e(1-c(g))} =\frac{\mathbb{1}_{\mathcal{D}}(x, y)}{\int_{\mathcal{D}} g(x',y') \,\dif x'\,\dif y'},
\end{equation}
which ensures that the selection density is properly normalized
\begin{equation}
	\int_{[0,1]^2} s(x,y; g) g(x,y)\, \dif x\,\dif y = 1.
\end{equation}
Thus, exactly one discordant edge is chosen for rewiring.
For $c(g)= 1$ the weight function is not well defined, but we know from the microscopic process that in this case no more rewiring events occur.
Once an edge is removed, one of its nodes, say $x$, forms a new edge with a node $y$ according to the probability distribution given by \Cref{eq:rewiring_probability}.
In the graphon limit, the rewiring probability becomes
\begin{multline}
    p(x, y; g_t) := \\
    \frac{1}{Z(x; g_t)} (1-|\theta(x) - \theta(y)|)(1-g_t(x,y)),
\end{multline}
where 
\begin{equation}
    Z(x; g_t) := \int_0^1 (1-|\theta(x) - \theta(y)|)(1-g_t(x,y)) \, \dif y.
\end{equation}
The transition rate $T(x, y, x', y';g_t)$, which defines the rate at which an edge at $(x', y')$ is replaced by an edge $(x, y)$, is then given by
\begin{multline}
	T(x, y, x', y'; g_t) := \frac{\lambda }{2} s(x', y'; g_t)\, \big[\delta(y-y') p(y', x; g_t) + \\
    \delta(x-x') p(x', y;g_t)\big].
\end{multline}
The Dirac-delta terms $\delta(\cdot)$ enforce that one node from the dissolved edge $(x', y')$ is retained in the newly formed edge. Each of the two nodes is selected with equal probability.
\begin{widetext}
	By balancing the rates of edge dissolution and formation, we obtain the balance equation for $g_t$
	\begin{align}\label{eq:balance}
		\frac{\partial g_t}{\partial t}(x,y)
		 & = \int_{[0,1]^2} \Big(T(x,y, x', y';g_t) g_t(x',y')- T(x',y',x,y; g_t)g_t(x,y) \Big) \,\dif x' \, \dif y'                                                                   \\
		 & = \frac{\lambda}{2} \left[ \int_0^1 s(x', y;g_t) p(y, x;g_t)g_t(x',y) \,\dif x' + \int_0^1 s(x, y';g_t)p(x, y;g_t)g_t(x,y') \,\dif y' \right]                           \\
		 & \qquad - \frac{\lambda}{2} s(x,y;g_t) g_t(x,y) \underbrace{\left[ \int_0^1 p(y, x';g_t) \,\dif x' + \int_0^1 p(x, y';g_t) \,\dif y' \right]}_{=2} \nonumber                   \\
		 & = \frac{\lambda}{2} \left[ p(y,x;g_t) \int_0^1 s(x', y;g_t)g_t(x',y) \,\dif x' + p(x,y;g_t) \int_0^1 s(x, y';g_t) g_t(x,y') \,\dif y' \right] \\
      & \qquad - \lambda s(x,y;g_t) g_t(x,y) \nonumber \\
		 & = \frac{\lambda}{2e (1-c(g_t))} [p(y, x;g_t) d(y; g_t) + p(x,y;g_t) d(x; g_t)] - \frac{\lambda}{e(1-c(g_t))} \mathbb{1}_\mathcal{D}(x,y) g_t(x,y),
	\end{align}
	where 
    \begin{equation}\label{eq:d}
        d(x; g) := \int_0^1 \mathbb{1}_\mathcal{D}(x, y) g(x, y) \, \dif y
    \end{equation}
    is the \textit{discordant degree function} for any graphon $g$. 
\end{widetext}
In some cases (see \Cref{eq:continued_growth}), however, the PDE predicts increasing consensus even as \mbox{$c(g_t)\to 1$}.
This behavior is inconsistent with the microscopic dynamics, which reaches an absorbing state once $C_r=1$.
Therefore we define
\begin{align}
  \left. \frac{\partial g_t}{\partial t} \right|_{c=1} =0
.\end{align}

\paragraph*{Closure approximation.}
In order to obtain an approximation for the dynamics of the consensus measure, we need to capture the ``saturation'' effect that slows down the growth of the consensus
\begin{equation}
    c_t := c(g_t)
\end{equation}
as the concordant domain $\mathcal{C}$ saturates, making rewiring into $\mathcal{C}$ less likely.
This effect also occurs when the discordant domain contains many edges while the concordant domain is sparse.
In this case, the increased edge density in the discordant domain increases the probability that a rewiring event produces a concordant edge.

We approximate $g_t$ by a step function determined by the average local edge density within the two domains:
\begin{equation}\label{eq:ansatz}
	g_t(x,y) \approx \begin{cases}
	  \rho_{\text{dis}}(c_t) = \frac{e(1-c_t)}{|\mathcal{D}|}       & \text{if } (x,y) \in \mathcal{D}   \\
	  \rho_{\text{con}}(c_t) = \frac{ec_t}{|\mathcal{C}|} & \text{if } (x,y) \in \mathcal{C}.
	\end{cases}
\end{equation}
Here, $|\mathcal{C}|,|\mathcal{D}|\in [0,1]$ are the geometric areas of the concordant and discordant domains, respectively, satisfying $|\mathcal C| + |\mathcal D| = 1$.

The evolution of the consensus measure is obtained by integrating the balance equation over the concordant domain $\mathcal{C}$. Because the total edge density $e$ is conserved, any edge added to $\mathcal{C}$ must be removed from $\mathcal{D}$. Thus,
\begin{equation}
  \frac{\dif c_t}{\dif t} = \frac{1}{e}\int_{\mathcal{C}} \frac{\partial g_t}{\partial t}\,\dif x\,\dif y = -\frac{1}{e} \int_\mathcal{\mathcal{D}} \frac{\partial g_t}{\partial t} \,\dif x\,\dif y 
.\end{equation}
Integrating \Cref{eq:balance} over $\mathcal{D}$ and utilizing the symmetry of the integrand to combine the transition probability terms yields
\begin{align}
  \frac{\dif c_t}{\dif t} & = - \int_\mathcal{D} \frac{\lambda}{e^2(1-c)} \Big[p(x,y;g_t) d(x; g_t) \Big] \,\dif x\,\dif y                                                           \\
		& \quad + \int_\mathcal{D} \frac{\lambda}{e^2(1-c)} g_t(x,y) \,\dif x\,\dif y                                                                       \nonumber   \\
		& = \frac{\lambda}{e} -\frac{\lambda }{e^2(1-c)} \int_0^1 q(x, g_t)\, d(x; g_t)\,\dif x \label{eq:dc}
,\end{align}
where $q(x, g_t):=\int_0^1 \mathbb{1}_\mathcal{D}(x,y) p(x,y;g_t) \,\dif y$.
Using the ansatz from \Cref{eq:ansatz}, the discordant degree function (defined in~\Cref{eq:d}) simplifies to
\begin{align}
  d(x; g_t) &= \int_0^1 \mathbb{1}_\mathcal{D}(x,y) \rho_{\text{dis}}(c_t) \,\dif y\\
  &= \rho_{\text{dis}}(c_t) \ell_\mathcal{D}(x),
\end{align}
where 
\begin{equation}\label{eq:ell}
    \ell_\mathcal{D}(x) := \int_0^1 \mathbb{1}_\mathcal{D}(x,y) \,\dif y
\end{equation}
is the measure of the discordant slice for node~$x$.

The probability
\begin{equation}
	p(x,y;g_t) = \frac{W(x,y)(1 - g_t(x,y))}{Z(x, g_t)}
\end{equation}
for a dissolved edge at $x$ to rewire to $y$ depends on the preference $W(x,y) := 1 - |\theta(x) - \theta(y)|$ 
and the local saturation level $(1 - g_t(x,y))$.
With our ansatz from \Cref{eq:ansatz}, the normalization factor $Z(x; g_t)$ decomposes into
\begin{align}
  Z(x; g_t) & = \int_0^1 \mathbb{1}_\mathcal{D}(x,y) W(x,y)(1 - \rho_{\text{dis}}(c_t)) \,\dif y   \nonumber                   \\
	  & \quad + \int_0^1 \mathbb{1}_\mathcal{C}(x,y) W(x,y)(1 - \rho_{\text{con}}(c_t)) \,\dif y                \\
	  & = (1 - \rho_{\text{dis}}(c_t)) Z_\text{dis}(x) \nonumber\\
      & \quad + (1 - \rho_{\text{con}}(c_t)) Z_{\text{con}}(x),
\end{align}
where $Z_{\text{dis}}(x) := \int_0^1 \mathbb{1}_{\mathcal{D}}(x,y)W(x,y) \,\dif y$ and $Z_{\text{con}}(x):=\int_0^1 \mathbb{1}_{\mathcal{C}}(x,y)W(x,y) \,\dif y$ measure the total affinity for a node~$x$ toward discordant and concordant connections, respectively.

For the probability $q(x;g_t)$ that a new edge formed by node $x$ lies in the discordant domain (see \Cref{eq:dc}),  we have
\begin{align}
  q(x; g_t) & = \int_0^1 \mathbb{1}_\mathcal{D}(x,y) p(x,y; g_t) \,\dif y                                                 \\
		& = \frac{(1 - \rho_{\text{dis}}(c_t)) Z_{\text{dis}}(x)}{(1 - \rho_{\text{dis}}(c_t)) Z_{\text{dis}}(x) + (1 - \rho_{\text{con}}(c_t)) Z_{\text{con}}(x)} \nonumber\\
		& = \left[1+ \frac{1-\rho_{\text{con}}(c_t)}{1-\rho_{\text{dis}}(c_t)} \cdot\frac{Z_{\text{con}}(x)}{Z_{\text{dis}}(x)} \right]^{-1} \\
        & = \left[1 + \alpha(c_t)R(x)\right]^{-1} ,
\end{align}
where 
\begin{equation}
    R(x) := \frac{Z_{\text{con}}(x)}{Z_{\text{dis}}(x)} > 0
\end{equation}
is the static affinity ratio and 
\begin{equation}
    \alpha(c) := \frac{1-\rho_{\text{con}}(c)}{1-\rho_{\text{dis}}(c)}
\end{equation}
refers to the saturation imbalance.

\begin{figure}
	\centering
	\includegraphics[width=.7\linewidth]{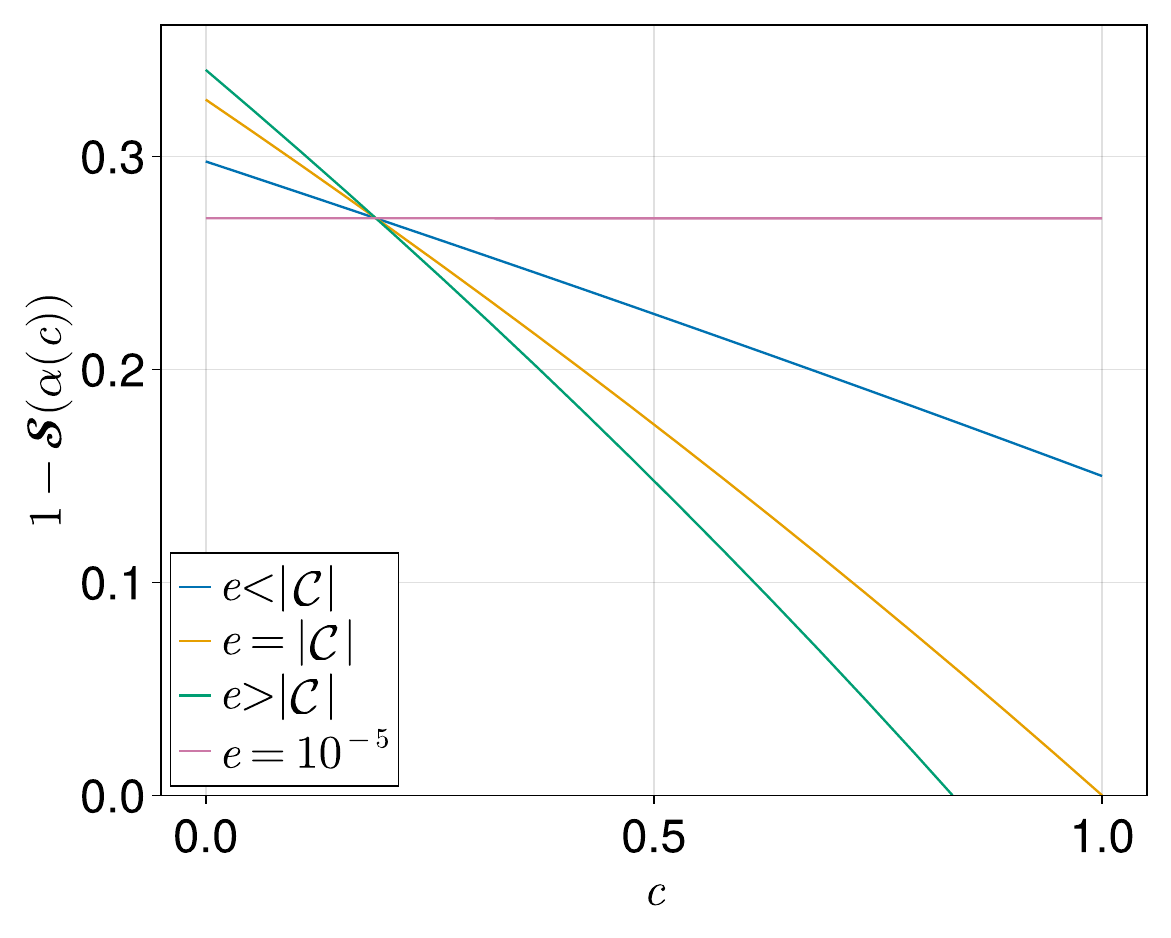}
	\caption{Illustration of the right-hand side of the ODE in~\cref{eq:appendix_ode} for the different regimes of the graphon ODE for $\lambda = 1$, $r=0.1$, uniform opinion distribution for different edge densities~$e$. When $e<|\mathcal{C}|$ the ODE is discontinuous at $c=1$, if $e>|\mathcal{C}|$, the ODE has a fixed point for $c<1$ and for very sparse graphons $e\ll1$, the ODE becomes asymptotically linear.}
	\label{fig:graphon_regimes}
\end{figure}

Putting everything together, \Cref{eq:dc} takes the closed-form
\begin{equation}\label{eq:appendix_ode}
  \frac{\dif c_t}{\dif t}  = \frac{\lambda}{e} \left[ 1 - \mathcal{S}(\alpha(c_t)) \right], 
\end{equation}
where the stagnation factor illustrated in~\Cref{fig:graphon_regimes} is given by
\begin{equation}
  \mathcal{S}(\alpha) := \frac{1}{|\mathcal{D}|} \int_0^1 \frac{\ell_\mathcal{D}(x)}{1+R(x)\, \alpha} \, \dif x.
\end{equation}

We distinguish between two regimes: the sparse regime, where the entire edge density can be transported into the concordant domain, i.e., $e < |\mathcal{C}|$, and the dense regime, where $e\ge |\mathcal{C}|$. For simplicity, we assume that $c=0$ is an admissible configuration.

In the sparse regime, \Cref{eq:appendix_ode} is valid for $c \in [0,1)$ and hence $\lim_{c \rightarrow 1} \alpha(c) = 1-\frac{e}{|\mathcal{C}|} > 0$, which implies $1+\alpha(1)R(x) > 1$. Therefore,
\begin{equation}\label{eq:continued_growth}
  \mathcal{S}(\alpha(1)) < \frac{1}{|\mathcal{D}|} \int_0^1 \ell_\mathcal{D}(x)\,\dif x = 1,
\end{equation}
so the ODE predicts continued growth of $c_t$ beyond $c=1$. To remain consistent with the microscopic dynamics, we impose the absorbing condition
\begin{equation}
  \left. \frac{\dif c_t}{\dif t}\right|_{c=1} = 0
,\end{equation}
which introduces a discontinuity in $\frac{\dif c_t}{\dif t}$ at $c=1$ in the sparse regime.

In the dense regime, we have $c_\text{max} = \frac{|\mathcal{C}|}{e}$ and $\alpha(c_\text{max}) = 0$.
This yields the maximal stagnation factor
\begin{equation}
  \mathcal{S}(\alpha(c_\text{max})) =1,
\end{equation}
resulting in a regular fixed point of the consensus dynamics at $c=\frac{|\mathcal{C}|}{e}$, which is approached from below.

The projected dynamics are captured well by the ODE in \Cref{eq:appendix_ode}, in both the sparse and the dense regime, as can be seen in~\Cref{fig:graphon,fig:graphon_dense_regime} in the main text.

In the limit of very sparse graphs, $e \rightarrow 0$, we have $\lim_{e \to 0} \alpha(c) = 1$, and the growth rate converges to the constant
\begin{equation}
  \frac{\lambda}{e}[1-\mathcal{S}(1)].
\end{equation}
The different stagnation factors for $c=0$ in~\Cref{fig:graphon_regimes} highlight that a high edge density within the discordant domain can increase the probability that rewiring events produce concordant edges.

In general, the stagnation factor $\mathcal{S}(\alpha)$ does not admit a closed-form expression. 
Although such an expression exists for the uniform distribution, we evaluate the integral numerically at each time step of the ODE integration. 
For the numerical integration, we use a one-dimensional grid with $500$ points together with the trapezoidal rule.

\begin{figure}
	\centering
	\includegraphics[width=\linewidth]{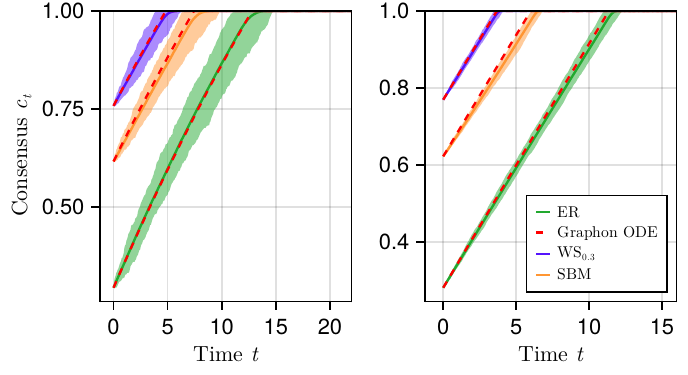}
	\caption{Evolution of the consensus measure $c_t$ compared to $\E[\rv{C}_r(t)]$ for different initial graph topologies. All parameters as in~\cref{fig:graphon} but normally distributed opinions $\theta_i \sim \mathcal{N}(0.5, 0.2)$. The normalized $L_1$ errors are less or equal then  $3 \times 10^{-3}$ (left) and  $5\times 10^{-3}$ (right).
	}
	\label{fig:graphon_appendix}
\end{figure}

Furthermore, we also evaluate the graphon approximation for a normal opinion distribution, $\theta_i \sim \mathcal{N}(0.5, 0.2)$. As shown in~\Cref{fig:graphon_appendix}, the derived ODE captures the consensus dynamics for normally distributed opinions accurately across the different initial graph topologies.

\newpage
\bibliography{ref}

\end{document}